\documentclass[twocolumn]{aastex62}

\newcommand{\HI}{\mbox{H\,{\sc i}}}
\newcommand{\HII}{\mbox{H\,{\sc ii}}}
\newcommand{\cow}{AT2018cow}
\newcommand{\cowhost}{CGCG~137-068}
\newcommand{\sii}{\mbox{[S\,{\sc ii}]}}

\newcommand{\nii}{\mbox{[N\,{\sc ii}]}}
\newcommand{\oiii}{\mbox{[O\,{\sc iii}]}}

\received{--}
\revised{--}
\accepted{--}
\submitjournal{ApJL}

\shorttitle{Molecular gas of CGCG~137-068}
\shortauthors{Morokuma-Matsui et al.}

\begin{document}

\title{ALMA observations of molecular gas in the host galaxy of \cow}

\correspondingauthor{Kana Morokuma-Matsui}
\email{kanamoro@ioa.s.u-tokyo.ac.jp}

\author[0000-0003-3932-0952]{Kana Morokuma-Matsui}
\altaffiliation{JSPS Fellow}
\affil{Institute of Astronomy, Graduate School of Science, The University of Tokyo, 2-21-1 Osawa, Mitaka, Tokyo 181-0015, Japan}

\author[0000-0001-7449-4814]{Tomoki Morokuma}
\affiliation{Institute of Astronomy, Graduate School of Science, The University of Tokyo, 2-21-1 Osawa, Mitaka, Tokyo 181-0015, Japan}

\author[0000-0001-8537-3153]{Nozomu Tominaga}
\affiliation{Department of Physics, Faculty of Science and Engineering, Konan University, 8-9-1 Okamoto, Kobe, Hyogo 658-8501, Japan}

\author[0000-0001-6469-8725]{Bunyo Hatsukade}
\affiliation{Institute of Astronomy, Graduate School of Science, The University of Tokyo, 2-21-1 Osawa, Mitaka, Tokyo 181-0015, Japan}

\author[0000-0002-9321-7406]{Masao Hayashi}
\affiliation{National Astronomical Observatory of Japan, 2-21-1 Osawa, Mitaka, Tokyo 181-8588, Japan}

\author[0000-0003-4807-8117]{Yoichi Tamura}
\affiliation{Department of Physics, Nagoya University, Furo-cho, Chikusa-ku, Nagoya 464-8602, Japan}

\author{Yuichi Matsuda}
\affiliation{National Astronomical Observatory of Japan, 2-21-1 Osawa, Mitaka, Tokyo 181-8588, Japan}
\affiliation{Graduate University for Advanced Studies (SOKENDAI), Osawa 2-21-1, Mitaka, Tokyo 181-8588}

\author{Kazuhito Motogi}
\affiliation{Graduate School of Sciences and Technology for Innovation, Yamaguchi University, Yoshida 1677-1, Yamaguchi, Yamaguchi 753-8512, Japan}

\author[0000-0002-8169-3579]{Kotaro Niinuma}
\affiliation{Graduate School of Sciences and Technology for Innovation, Yamaguchi University, Yoshida 1677-1, Yamaguchi, Yamaguchi 753-8512, Japan}

\author{Masahiro Konishi}
\affiliation{Institute of Astronomy, Graduate School of Science, The University of Tokyo, 2-21-1 Osawa, Mitaka, Tokyo 181-0015, Japan}



\begin{abstract}
We investigate the molecular gas in, and star-formation properties of, the host galaxy (\cowhost) of a mysterious transient, \cow, at kpc and larger scales, using archival band-3 data from the Atacama Large Millimeter/submillimeter Array (ALMA).
\cow~is the nearest Fast-Evolving Luminous Transient (FELT), and this is the very first study unveiling molecular-gas properties of FELTs.
The achieved rms and beam size are 0.21~mJy~beam$^{-1}$ at a velocity resolution of $40$~km~s$^{-1}$ and $3''.66\times2''.71$ ($1.1~{\rm kpc} \times 0.8~{\rm kpc}$), respectively.
CO($J$=1-0) emission is successfully detected.
The total molecular gas mass inferred from the CO data is $(1.85\pm0.04)\times10^8$~M$_\odot$ with the Milky Way CO-to-H$_2$ conversion factor.
The H$_2$ column density at the \cow~site is estimated to be $8.6\times10^{20}$~cm$^{-2}$.
The ALMA data reveal that
(1) \cowhost~is a normal star-forming (SF) dwarf galaxy in terms of its molecular gas and star-formation properties and 
(2) \cow~is located between a CO peak and a blue star cluster.
These properties suggest on-going star formation and favor the explosion of a massive star as the progenitor of \cow.
We also find that \cowhost~has a solar or super-solar metallicity.
If the metallicity of the other FELT hosts is not higher than average, then some property of SF dwarf galaxies other than metallicity may be related to FELTs.
\end{abstract}



\section{Introduction} \label{sec:intro}

Recent high-cadence and wide-field surveys have discovered a new class of objects: Fast-Evolving Luminous Transients (FELT, e.g., \citealt{Drout:2014aa,Rest:2018aa}; Tominaga et al.~submitted), providing various observed properties such as light curves, spectral energy distributions and environments.
The properties of FELTs which distinguish them from other classes of transients are their featureless spectra and quickly declining light curves ($t_{1/2}\sim$ several days).
\cow~is one of the FELTs, but its close proximity ($60$~Mpc) provides us a unique opportunity to study in detail physical properties of a FELT for the first time.
It was discovered by the ATLAS survey on June 16, 2018, in the disk of dwarf star-forming (SF) galaxy, \cowhost~\citep{Prentice:2018aa}, and since that time it has been monitored at various wavelengths from radio to $\gamma$-ray \citep[e.g.,][]{Margutti:2019aa,Ho:2019aa,Perley:2019aa}.
It is characterized by a rapid rise time in brightness (a few days), a blue spectrum ($\sim30,000$~K), high optical luminosity ($4\times10^{44}$~erg~s$^{-1}$), initially featureless optical spectra, no $\gamma$-ray flash, high radio luminosity, and long-lived mm-wavelength emission \citep{Prentice:2018aa,Margutti:2019aa,Ho:2019aa,Perley:2019aa,Kuin:2019aa}.
With these observational constraints, various scenarios for the event have been proposed, and are roughly classified into two groups: massive star explosions \citep{Perley:2019aa, Margutti:2019aa}, and tidal disruption events (TDEs) in which a white dwarf is torn apart by an intermediate-mass black hole \citep{Kuin:2019aa,Perley:2019aa}.

Insight into the nature of transients is provided not only by observing the objects themselves but also 
their host galaxies, especially in order to distinguish whether events are related to the deaths of massive stars (i.e., on-going star formation) or not.
\cite{Roychowdhury:2019aa} found that \cow~resides in a ring-like \HI~structure seen in $\sim6$-arcsec resolution data taken  with the Giant Metrewave Radio Telescope (GMRT).
They concluded that the \HI~data indicates a massive-star scenario for \cow, since such a ring is an ideal site for active star formation.
In contrast, \cite{Michaowski:2019aa} reported an absence of atomic gas at the \cow~site with a $\sim14$-arcsec beam, and claimed that a TDE is a more plausible scenario for the progenitor of \cow~based on their GMRT observations.
The cold gas properties of the \cow~site are still controversial, and it is important to investigate it using molecular gas, the raw material of star formation.

In this paper, we study the molecular-gas and star-formation properties of \cowhost~and local \cow~site using CO($J$=1-0) data taken with the Atacama Large Millimeter/submillimeter Array (ALMA).
We adopt a distance to \cowhost~of $60$~Mpc (1$''\sim291$~pc) and cosmology parameters of $(h, \Omega_{M}, \Omega_\Lambda)=(0.7, 0.3, 0.7)$ throughout the paper.

\section{Data and Analysis} \label{sec:data}

ALMA band-3 Time-Division-Modes (TDM) data were obtained on June 30 and July 16, 2018, with an antenna configuration of C43-1 in a target-of-opportunity (ToO) observation (project code of 2017.A.00045.T, PI: Steve Schulze).
Antenna baselines range from 15~m to 161~m, and a maximum recoverable scale is $29$ arcsec or 8.4~kpc.
Since the original purpose of this observation is to measure the continuum flux of \cow~at band 3 ($100.00-115.99$~GHz), the data are taken in TDM mode, i.e., a frequency resolution of $15.625$~MHz, corresponding to a velocity resolution of $40.64$~km~s$^{-1}$ at the observing frequency.
The on-source integration time of each execution was $19.7$~min.
The CO($J$=1-0) emission line (rest frequency $\nu_{\rm rest}$ of 115.271202~GHz) was covered in one of the upper-sideband spectral windows (a CO spectrum is shown in Figure~\ref{fig:alma}).

Data calibration and imaging were conducted with the standard ALMA data analysis package, the Common Astronomy Software Applications \citep[CASA,][]{McMullin:2007aa,Petry:2012aa}.
The absolute flux and gain fluctuations of the 12-m data were calibrated with the ALMA Science Pipeline (version of r40896 of Pipeline-CASA51-P2-B) in the CASA 5.1.1 package.
The flux accuracy of the ALMA 12-m band-3 data is reported to be better than 5~\% (ALMA proposer's guide).
Continuum emission is subtracted using the {\tt UVCONTSUB} task and a $^{12}$CO($J$=1-0) mosaic data cube is generated with the {\tt TCLEAN} task in CASA version 5.4 with options of {\tt Briggs} weighting with a {\tt robust} parameter of 0.5, {\tt auto-multithresh} mask with standard values for 12-m array data provided in the CASA Guides for auto-masking\footnote{\url{https://casaguides.nrao.edu/index.php/Automasking_Guide}} ({\tt sidelobethreshold} of 2.0, {\tt noisethreshold} of 4.25, {\tt minbeamfrac} of 0.3, {\tt lownoisethreshold} of 1.5, and {\tt negativethreshold} of 15.0), and {\tt niter} of 10000.
The achieved synthesized beam is $3''.66\times2''.71$ ($1.1~{\rm kpc} \times 0.8~{\rm kpc}$ at the distance of \cowhost) with a P.A. of $-27.8^\circ$.
The achieved rms noise ($\sigma_{\rm rms}$) is 0.21~mJy~beam$^{-1}$ after primary beam correction at the velocity resolution of 40.64~km~s$^{-1}$ in the box area centered on the galaxy center with a width and height of 25~arcsec in the emission-free channels (Figure~\ref{fig:alma}b).

\section{Properties of an \cowhost}\label{sec:results}
In this section, we present local-site (Section~\ref{sec:local}) and host-galaxy properties of \cow~derived with the ALMA data (Section~\ref{sec:integrated}).

\subsection{1-kpc-scale properties} \label{sec:local}

\begin{figure*}[t]
\begin{center}
\includegraphics[width=180mm, bb=0 0 1198 637]{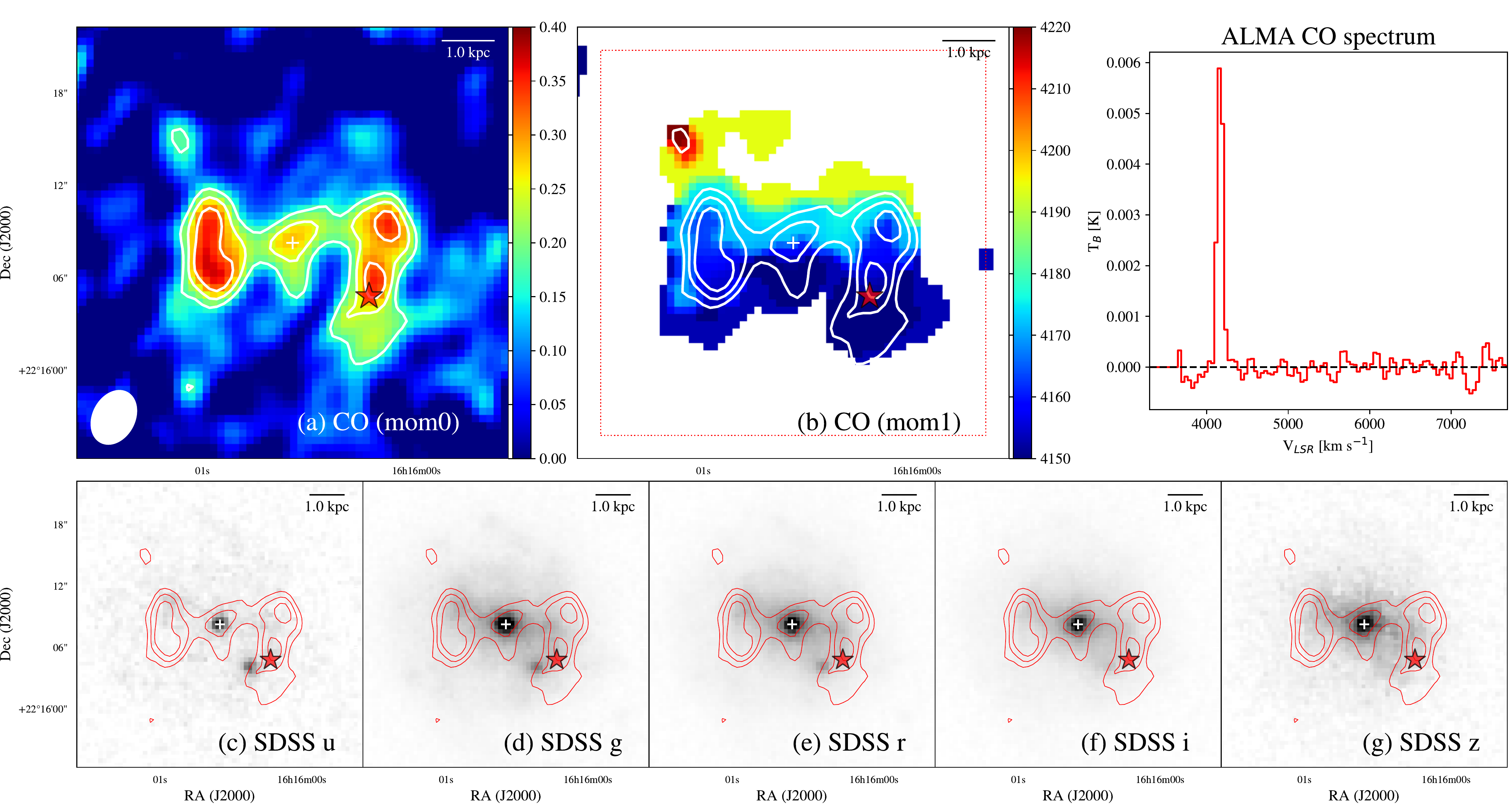}
\end{center}
\caption{(a) ALMA CO($J$=1-0) integrated intensity map in units of Jy~beam$^{-1}$~km~s$^{-1}$ and (b) velocity field of \cowhost~in units of km~s$^{-1}$ (``radio''-definition LSR velocity).
SDSS optical images are also shown for a comparison: (c) $u$, (d) $g$, (e) $r$, (f) $i$, (g) $z$.
Contours in each figure indicate the CO($J$=1-0) integrated intensity with contour levels of $(2.5, 3.5, 4.5)\times\sigma_{\rm rms, I.I.}$, where $\sigma_{\rm rms, I.I.}$ is $0.07$~Jy~beam$^{-1}$~km~s$^{-1}$.
Positions of the \cow~and galaxy center are indicated with a red star and a white cross, respectively.
The ALMA beam ($3''.66\times2''.71$ with a position angle of $-27.8^\circ$) is indicated as a while-filled circle at the bottom-left corner of the panel (a).
The ALMA CO spectrum averaged over the area enclosed by red dotted line in the panel (b) is also shown at the upper right corner.
}
\label{fig:alma}
\end{figure*}

Figure~\ref{fig:alma} shows ALMA CO($J$=1-0) moment maps
and optical images obtained in the Sloan Digital Sky Survey (SDSS) of \cowhost.
The basic and derived parameters of \cow~site are summarized in Table~\ref{tab:local}.
The ALMA data reveal that the \cow~is located at the edge of one of the CO peaks to the west of the galaxy center.
The H$_2$ column density, $N({\rm H}_2)$, at the \cow~site is estimated to be $8.6\times10^{20}$~cm$^{-2}$, using the standard CO-to-H$_2$ conversion factor from the Milky-Way of $X_{\rm CO}=2.0\times 10^{20}$~cm$^{-2}$~(K~km~s$^{-1}$)$^{-1}$ \citep{Bolatto:2013rr}\footnote{
The conversion from Jy~beam$^{-1}$ to Kelvin is done based on the Rayleigh-Jeans approximation with an equation of $T=1.222\times10^6\frac{I}{\nu^2 \theta_{maj}\theta_{min}}$, where $T$ is the brightness temperature in Kelvin, $\nu$ is the observing frequency in GHz, $\theta_{maj}$ and $\theta_{min}$ are half-power beam widths along the major and minor axes in arcsec, respectively and $I$ is the brightness in Jy~beam$^{-1}$.}
since \cowhost~has a solar or super-solar metallicity (see below).
This corresponds to a molecular gas mass surface density, $\Sigma_{\rm mol}$, of $14$~M$_\odot$~pc$^{-2}$ (Figure~\ref{fig:alma}).  
The CO velocity field is qualitatively consistent with the \HI~velocity field, showing a monotonic velocity increase from south to north.  
No disturbed velocity field is observed at the \cow~site with a velocity resolution of $\sim40$~km~s$^{-1}$.

\begin{table}
\begin{center}
\caption{Basic properties of \cow~site and its host}
  \begin{tabular}{lcl}
      \hline
      \hline
      Parameter & Value & Ref. \\ 
      \hline
      \multicolumn{3}{l}{\underline{\it Local site (\cow)}}\\
      R.A. & $16^{h}16^{m}00^{s}.2242$ & 1\\ 
      Dec. & $+22^{\circ}16'04''.890$ & 1\\ 
      $R_{\rm \cow}^{*}$ & $6''.0$ (1.75~kpc) & --\\ 
      $\Sigma_{\rm mol}$ (M$_\odot$~pc$^{-2}$) & $14$ & this study\\ 
      $N({\rm H}_2)$ (cm$^{-2}$) & $8.6\times10^{20}$ & this study\\ 
      \hline
      \multicolumn{3}{l}{\underline{\it Host galaxy (\cowhost)}}\\
      R.A. & $16^{h}16^{m}00^{s}.57$ & SDSS\\ 
      Dec. & $+22^{\circ}16'08''.24$ & SDSS\\ 
      Redshift & 0.0141 & SDSS\\ 
      Distance (Mpc) & 60 & 2\\ 
      M$_{\rm star}$ (M$_\odot$) & $10^{9.15}$ & 2\\ 
      SFR (M$_\odot$~yr$^{-1}$) & $0.22$ & 2\\
      sSFR (yr$^{-1}$) & $1.56\times10^{-10}$ & 2\\ 
      12+log(O/H)$_{\rm MPA-JHU}$ & $8.96$ & SDSS\\ 
      M$_{\rm atom}$ (M$_\odot$) & $(6.6\pm0.9)\times10^8$ & 3\\ 
      $I_{\rm CO}$ (K~km~s$^{-1}$) & $0.575\pm0.020$ & this study\\
      $L'_{\rm CO}$ (K~km~s$^{-1}$~pc$^{2}$) & $(4.34\pm0.15)\times 10^7$ & this study\\
      FWHM$_{\rm CO}^{***}$ (km~s$^{-1}$) & $85.5$ & this study\\
      M$_{\rm mol}$ (M$_\odot$) & $(1.85\pm0.04)\times10^8$ & this study\\ 
      SFE(mol) (yr$^{-1}$) & $1.2\times10^{-9}$ & this study\\ 
      SFE(gas)$^{**}$ (yr$^{-1}$) & $2.6\times10^{-10}$ & this study\\ 
      M$_{\rm atom}$/M$_{\rm star}$ & $0.47$ & 3\\ 
      M$_{\rm mol}$/M$_{\rm star}$ & $0.13$ & this study\\ 
      M$_{\rm mol}$/M$_{\rm atom}$ & $0.29$ & this study\\ 
      \hline
    \end{tabular}\label{tab:local}
\end{center}
\tablecomments{
$^*$Galactocentric distance.
$^{**}$SFE(gas) = SFR/(M$_{\rm atom}$+M$_{\rm mol}$).
$^{***}$From Gaussian fitted curve.
Ref. 1: \cite{Bietenholz:2018aa},
2: \cite{Perley:2019aa},
3: \cite{Roychowdhury:2019aa}
}
\end{table}

\subsection{Galactic-scale properties} \label{sec:integrated}

The properties related to molecular gas are presented in Table~\ref{tab:local}, together with other basic properties.
The total molecular gas mass is estimated to be $(1.85 \pm 0.04)\times 10^8$~M$_\odot$.
Both the stellar mass ($M_{\rm star}$) and star formation rate (SFR) values are taken from \cite{Perley:2019aa}, the gas-phase metallicity is measured from the SDSS spectrum, and the atomic gas mass ($M_{\rm atom}$) is from \cite{Roychowdhury:2019aa}.
The location of the host galaxy in the Baldwin, Phillips \& Telervich (BPT) diagram suggests that it is a SF galaxy \citep[$\log{(\nii/{\rm H\alpha})}$=$-0.35$, $\log{(\oiii/{\rm H\alpha})}$=$-0.40$, $\log{(\sii/{\rm H\alpha})}$=$-0.43$,][]{Baldwin:1981aa}.
Note that the metallicities of \cowhost~estimated with different calibrations are nearly solar or super-solar \citep[cf., 8.69 for the Solar,][]{Asplund:2009fh}:
8.96 \citep{Tremonti:2004rf} with ``MPA-JHU'' calibration\footnote{\url{https://wwwmpa.mpa-garching.mpg.de/SDSS/}};
9.35 \citep{Zaritsky:1994zt} and 8.98 \citep{Pilyugin:2005os} with ``$R_{23}$'';
8.77 \citep{Denicolo:2002aa} and 8.65 \citep{Pettini:2004jl} with ``N2'';
8.71 with ``O3N2'' \citep{Pettini:2004jl}.

\begin{figure*}[t]
\begin{center}
\includegraphics[width=180mm, bb=0 0 1400 931]{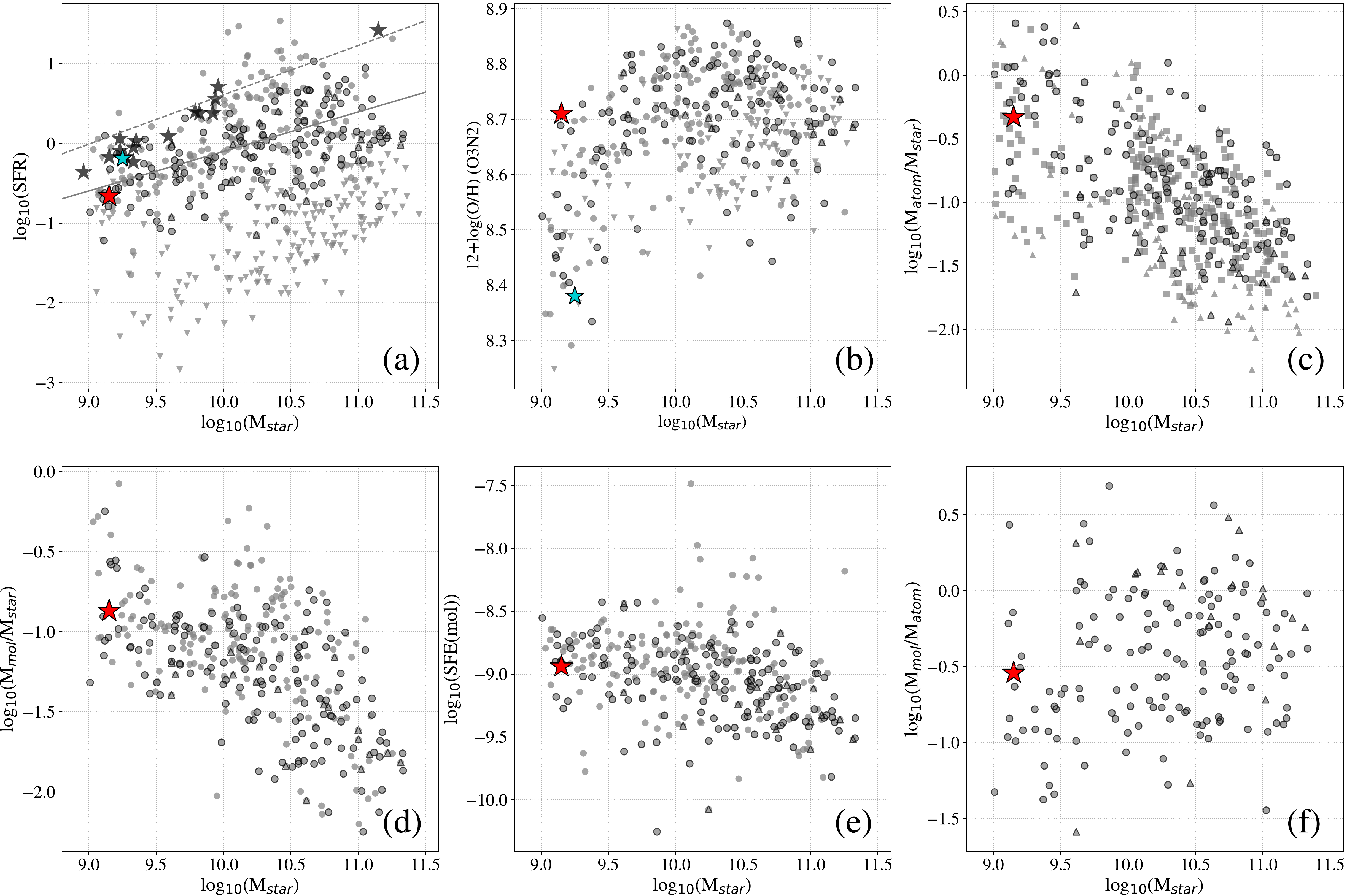}
\end{center}
\caption{Basic properties of \cowhost~\citep[red-filled star, assuming IMF of][]{Chabrier:2003oe} as a function of $M_{\rm star}$:
(a) $\log{({\rm SFR})}$,
(b) Matallicity, $12+\log{({\rm O/H})}$
(c) $\log{(M_{\rm atom}/M_{\rm star})}$
(d) $\log{(M_{\rm mol}/M_{\rm star})}$
(e) $\log{({\rm SFE(mol)})}=\log{({\rm SFR}/M_{\rm mol})}$
(f) $\log{(M_{\rm mol}/M_{\rm atom})}$.
For a reference, xCOLD GASS \citep{Saintonge:2017aa} and xGASS \citep{Catinella:2018aa} galaxies are plotted (grey-filled circle: CO detected, grey-filled down-pointing triangle: CO non-detected xCOLD GASS galaxies, grey-filled square: \HI~detected, grey-filled triangle: \HI~tentatively detected xGASS galaxies, black-open circle: both CO and \HI~detected, black-open triangle: CO detected and \HI~tentatively detected xCOLD GASS galaxies), assuming \cite{Chabrier:2003oe} IMF.
Other FELTs (redshift of $0.12<z<1.56$, a median redshift of 0.485) from \cite{Pursiainen:2018aa} also indicated as black-filled stars \citep[not mentioned in the article but probably assuming IMF of][]{Rana:1992aa}.
A FELT host in the \cite{Pursiainen:2018aa} sample with SDSS spectrum is also plotted adopting MPA-JHU data (blue-filled star).
SF main sequences at $z=0.01414$ (a redshift of \cow) and $z=0.485$ (a median redshift of the other FELTs) are presented as grey solid and dashed lines, respectively \citep{Speagle:2014aa}.
}
\label{fig:intg_prop}
\end{figure*}

In Figure~\ref{fig:intg_prop}, we compare the integrated galactic properties (SFR, metallicity, $M_{\rm atom}/M_{\rm star}$, $M_{\rm mol}/M_{\rm star}$, star formation efficiency of molecular gas (SFE(mol)), and $M_{\rm mol}/M_{\rm atom}$, as a function of $M_{\rm star}$) of \cowhost~with those of xGASS \citep{Catinella:2018aa} and xCOLD-GASS galaxies \citep{Saintonge:2017aa} at similar redshifts.  
Note that SFE(mol) of \cowhost~is comparable to that of host galaxies of long-duration gamma-ray bursts \citep[GRBs, e.g.,][]{Hatsukade:2019aa}, which are thought to be associated with the explosion of massive stars.  
Overall, \cowhost~is a normal low-mass SF galaxy in terms of molecular-gas and star-formation properties, but has a relatively high metallicity considering its low stellar mass.

\section{Implications for \cow~progenitor}\label{sec:comparison}

\begin{table*}
\begin{center}
\caption{Scoring table for progenitor models of \cow}
  \begin{tabular}{|ll|ccc|cccc|}
      \hline
      &&\multicolumn{3}{c}{\underline{\it \cow~site}} & \multicolumn{4}{|c|}{\underline{\it \cowhost}}\\
      \multicolumn{2}{|c|}{Scenario} & $R_{\cow}$ & $N({\rm H}_2)$ & \HII & BPT & SF activity & morphology & metallicity\\ 
      \hline
      \multicolumn{2}{|l|}{TDEs} & $\times$ & -- & -- & $\times$ & $\times$  & -- & -- \\ 
      CCSNe & SNe II & $\checkmark$ & $\times$ & Maybe $\times$ & $\times$ & $\checkmark$ & $\checkmark$ & Maybe $\times$ \\ 
      & SNe Ib & $\checkmark$ & $\checkmark$ & Maybe $\checkmark$ & $\checkmark$ & $\checkmark$ & $\checkmark$ & Maybe $\checkmark$ \\ 
      & SNe Ic & $\checkmark$ & $\checkmark$ & Maybe $\checkmark$ & $\checkmark$ & $\checkmark$ & $\times$ & Maybe $\checkmark$ \\ 
      & SNe Ic-BL w/ GRB & $\checkmark$ & $\checkmark$ & Maybe $\checkmark$ & $\checkmark$ & $\checkmark$ & $\checkmark$ & $\times$ \\ 
      & SNe Ic-BL w/o GRB & $\checkmark$ & $\checkmark$ & Maybe $\checkmark$ & $\checkmark$ & $\checkmark$ & $\checkmark$ & $\checkmark$ \\ 
      \hline
    \end{tabular}\label{tab:marubatsu}
\end{center}
\end{table*}

The ALMA data revealed that \cow~is located at the edge of one of the CO peaks to the west side of the galaxy center, and its host is a normal low-mass SF galaxy in terms of star-formation and cold-gas properties, except for its high metallicity.
These properties of \cow~site and \cowhost, i.e., abundant material for future star formation and proximity to a young star cluster, are suggestive of the progenitor scenarios featuring the explosion of a massive star.  
In this section, we first compare local-site and host-galaxy properties of \cow~with those of CCSNe and TDEs, then further compare with subclasses of CCSNe.  
The comparison is summarized in Table~\ref{tab:marubatsu}.  
In addition, we also mention other FELTs, comparing the properties of their host galaxies to the properties of \cowhost.

First, in terms of location within a host galaxy, core-collapse SNe (CCSNe) tend to reside in the outskirts of galaxies \citep{Galbany:2017aa}, whereas TDEs are found generally at galactic centers \citep{Komossa:1999aa,Gezari:2012aa,Miller:2015aa}.  
\cow~is located at $\sim$6~arcsec (1.75~kpc) from the galaxy center, which favors the CCSNe scenario.  
However, it should be noted that an off-center TDE event (12.5~kpc from center) was recently detected in X-rays in a large lenticular galaxy \citep{Lin:2018aa}.  
This event is considered to be a TDE by a few $10^4$~M$_\odot$ black hole.

Second, it is claimed that transients of different types may arise preferentially in certain types of galaxies.  
TDEs are claimed to reside primarily in quiescent Balmer-strong galaxies (i.e., post-starburst galaxies) unlike \cowhost~\citep[e.g.,][]{Arcavi:2014aa,French:2016aa}.  
In addition, TDE hosts tend to have a LINER/Seyfert nucleus based on BPT diagnostics, similar to those in other quiescent Balmer-strong galaxies \citep{French:2017aa}.  
Statistical studies of CCSNe host galaxies, on the other hand, show that the number ratio of Type-I (H-poor, Ib and Ic) to Type-II (H-rich) CCSNe is not a strong function of the Hubble type \citep{Hakobyan:2014aa}, 
that the Type-I-to-Type-II CCSNe number ratio tends to be high for SF galaxies based on BPT diagnostics \citep{Hakobyan:2014aa},
and that Type-I CCSNe found in dwarf galaxies are either SNe~Ib or broad-line SNe~Ic (SNe~Ic-BL), whereas normal SNe~Ic dominate in giant galaxies \citep{Arcavi:2010aa}.  
Thus, the galaxy type of \cowhost~also suggests a CCSN scenario, especially SNe~Ib or SNe~Ic-BL.

\begin{figure}[t]
\begin{center}
\includegraphics[width=80mm, bb=0 0 497 496]{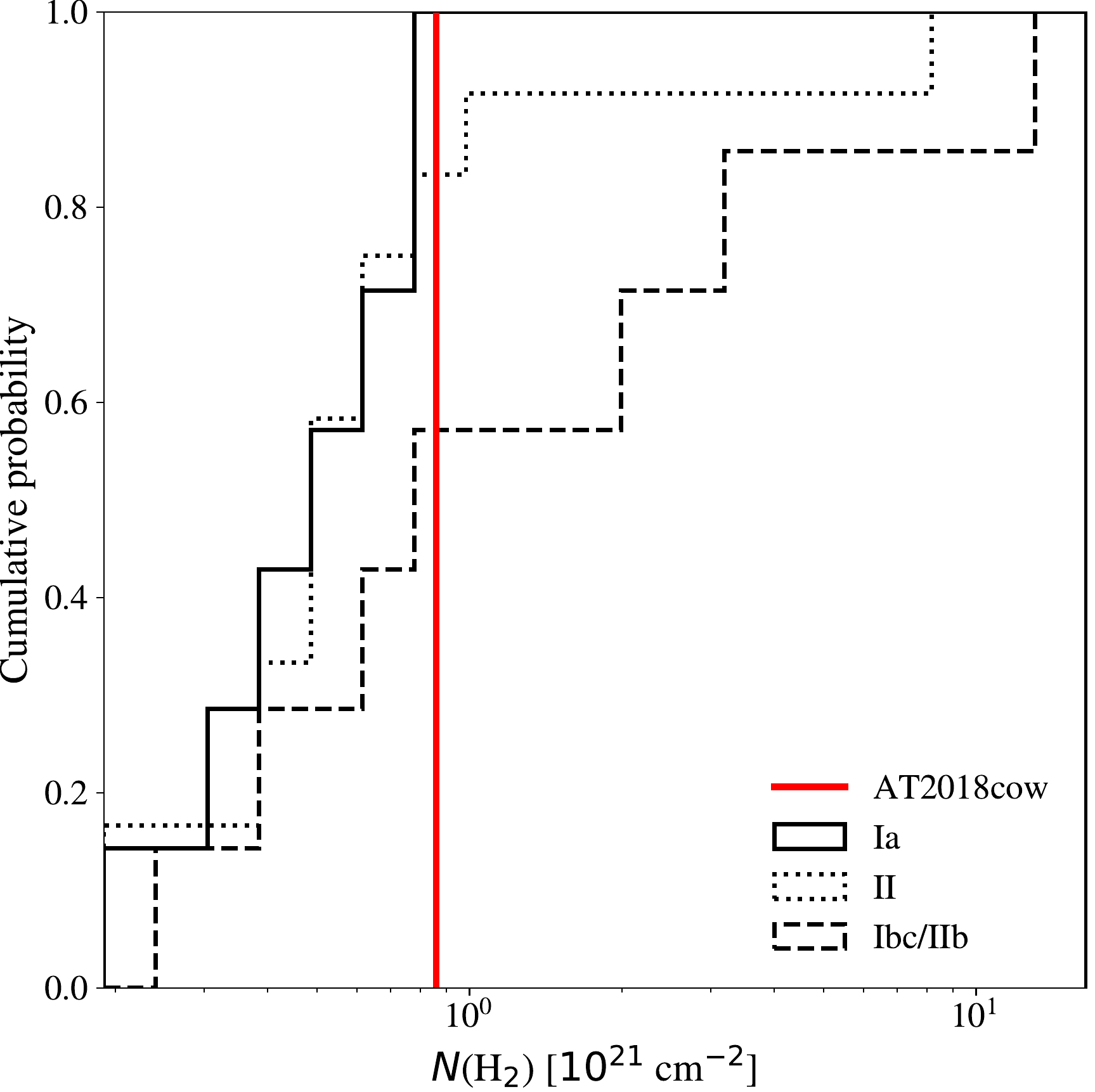}
\end{center}
\caption{Comparison between H$_2$ column density at the \cow~site and that of other SNe sites from \cite{Galbany:2017aa}.
Note that most of the SNe Ia and II data are upper limits.
}
\label{fig:nh2}
\end{figure}

Third, it is claimed that the kpc-scale $N({\rm H}_2)$ tends to be higher at Type-I CCSNe sites than Type-II CCSNe sites in nearby galaxies \citep{Galbany:2017aa}.  
Compared to these SNe, $N({\rm H}_2)$ at the \cow~site is slightly higher than the Type-II CCSNe (mostly upper limits in the previous study), but comparable to Type-I CCSNe, which have a median of $N({\rm H}_2)\sim9.4\times10^{20}$~cm$^{-2}$ (Figure~\ref{fig:nh2}).  
Note that \HI~emission is found to be not so strong or absent at the \cow~site \citep{Roychowdhury:2019aa,Michaowski:2019aa}.  
This is consistent with the previous studies on cold gas properties of nearby galaxies, where the dominant phase changes from atomic to molecular gas at $\Sigma_{\rm mol}\sim10$~M$_\odot$~pc$^{-2}$ \citep{Bigiel:2008rt}.

Fourth, the association between CCSNe and \HII~regions is claimed to be different according to types of SNe: a higher association is found for Type-I CCSNe than Type-II CCSNe, suggesting that the progenitors of Type-I CCSNe are more massive than Type-II CCSNe \citep{Anderson:2008aa,Crowther:2013aa}.  
In the bottom of Figure~\ref{fig:alma}, \cow~seems to be located between a $u$-band bright cluster and the molecular gas peak.  
The cluster tends to be brighter at shorter wavelengths, suggesting it is young.  
The distance between the \cow~site and the cluster is $640$~pc, which is much larger than the typical \HII~region but comparable to supergiant \HII~regions such as NGC~5461 of M~101 \citep[a radius of $\sim500$~pc,][]{Crowther:2013aa}.  
Although we do not have H$\alpha$~data, the proximity between the \cow~and the nearby young stellar cluster may suggest the \cow~site resembles local environments of Type-I CCSNe.

Finally, the high metallicity of \cowhost~is one of its more noteworthy properties.  
Generally, it is reported that the metallicity at explosion sites is not a strong function of CCSNe types \citep{Kuncarayakti:2018aa}, though the sites of Type-I SNe may have a slightly higher metallicity than those of Type-II SNe \citep{Anderson:2010aa}.  
Some observations suggest that metallicity at the site of SNe~Ic-BL without a long GRB tends to be higher than that of SNe~Ic-BL with a long GRB \citep[e.g.,][]{Modjaz:2008aa,Japelj:2018aa}, although the sample size is small and the statistical significance is low \citep[see also][]{Modjaz:2019aa}.  
If the difference is real, these two populations are likely to be intrinsically different, possibly in terms of success or failure of break-out of a jet \citep{Lazzati:2012aa,Japelj:2018aa}.
Considering that \cowhost~was not accompanied by a GRB, the observed properties of \cowhost~resemble those of SNe~Ic-BL without a GRB.

In Figure~\ref{fig:intg_prop}a, we plot other FELTs listed as the ``Gold sample'', i.e., most secure sample, with stellar mass and SFR estimations in \cite{Pursiainen:2018aa}.  
We can see that these host galaxies are also SF galaxies at their redshifts ($0.12<z<1.56$, a median redshift of 0.485)\footnote{The ``SF main sequence'' evolves with time \citep{Speagle:2014aa}.} and tend to have low stellar mass ($8.26<\log{({\rm M_{star}}/{\rm M}_\odot)}<11.15$, a median of $9.33$), as \cowhost.
One of the FELT hosts in \cite{Pursiainen:2018aa} without SFR estimation has SDSS spectroscopic data.
In Figures~\ref{fig:intg_prop}a and b, we plotted this host using SDSS data, showing that it is a low-mass SF galaxies with a relatively low metallicity.
Meanwhile, some metallicity calibrations suggest that it has a super-solar metallicity\footnote{
8.60 \citep{Tremonti:2004rf} with ``MPA-JHU'' calibration;
8.82 \citep{McGaugh:1991ee}, 8.93 \citep{Zaritsky:1994zt} and 8.45 \citep{Pilyugin:2005os} with ``$R_{23}$'';
8.51 \citep{Denicolo:2002aa} and 8.36 \citep{Pettini:2004jl} with ``N2'';
8.38 with ``O3N2'' \citep{Pettini:2004jl}}.
This suggests that some properties of dwarf SF galaxies other than metallicity may be related to FELTs.
However, the sample size ($N=2$) is too small to conclude it and metallicity measurements of other FELT hosts are required.

\section{Summary} \label{sec:summary}

We investigate the molecular-gas and star-formation properties of the host galaxy (\cowhost) and local site of \cow~using archival ALMA CO($J$=1-0) data.
We found that:
\begin{enumerate}
\item \cow~is located between one of the CO peaks and a blue stellar cluster, both of which are indicators of on-going star formation.

\item The H$_2$ column density of the \cow~site is $8.6\times10^{20}$~cm$^{-2}$, which is slightly higher than that of Type-II CCSNe sites and comparable to that of Type-I CCSNe sites. 

\item The total molecular gas mass is $(1.85\pm0.04)\times10^8$~M$_\odot$, using a Milky-Way CO-to-H$_2$ conversion factor.
With the literature data, $M_{\rm mol}/M_{\rm star}$, SFE(mol), and $M_{\rm mol}/M_{\rm atom}$ of \cowhost~are $0.13$, $1.2\times10^{-9}$~yr$^{-1}$, and 0.29, respectively.

\item Compared to reference galaxies at similar redshift (xGASS and xCOLD GASS galaxies), \cowhost~is a normal SF dwarf galaxy in terms of SFR, $M_{\rm atom}/M_{\rm star}$, $M_{\rm mol}/M_{\rm star}$, SFE(mol), and $M_{\rm mol}/M_{\rm atom}$.
The gas-phase metallicity of \cowhost~is relatively high (solar or super-solar) for the stellar mass of \cowhost.

\item Compared to the previous studies on known transients, both the host-galaxy and local-site properties of the \cow~are indicative of massive-star explosions, especially SNe Ic-BL without GRB, although the observed properties of \cow~itself cannot be fully explained by SNe~Ic-BL models.

\end{enumerate}

For a more detailed understanding of \cow, deep integral-field-spectroscopy observations of \cowhost~are necessary to investigate the H$\alpha$ distribution and metallicity around the \cow~site, and to determine the age of the nearby young cluster.
In addition, follow-up observations of the other FELT hosts, such as metallicity measurements, are required to place \cow~among the general population of FELT hosts.
If the metallicity of the other FELT hosts is not higher than average, then some property of dwarf SF galaxies other than metallicity may be related to FELTs.




\acknowledgments
We are grateful to the anonymous referee for his/her helpful comments and to the PDJ collaboration for providing opportunities for fruitful discussions.
KMM thank Dr. Yuu Niino for useful discussions.
This study is supported by JSPS KAKENHI grant No.s of 15H02075, 15H00784, 16H02158, 17H06130, 17KK0098, 17H04831, 18H03721, and 19K03925.
We also appreciate the kind help of Prof. Michael W. Richmond in improving the English grammar of the manuscript.

This paper makes use of the following ALMA data: ADS/JAO.ALMA\#2017.A.00045.T.
ALMA is a partnership of ESO (representing its member states), NSF (USA) and NINS (Japan), 
together with NRC (Canada), MOST and ASIAA (Taiwan), and KASI (Republic of Korea), in 
cooperation with the Republic of Chile. The Joint ALMA Observatory is operated by 
ESO, AUI/NRAO and NAOJ.


Funding for the Sloan Digital Sky Survey IV has been provided by the Alfred P. Sloan Foundation, the U.S. Department of Energy Office of Science, and the Participating Institutions. SDSS-IV acknowledges
support and resources from the Center for High-Performance Computing at
the University of Utah. The SDSS web site is \url{www.sdss.org}.

SDSS-IV is managed by the Astrophysical Research Consortium for the 
Participating Institutions of the SDSS Collaboration including the 
Brazilian Participation Group, the Carnegie Institution for Science, 
Carnegie Mellon University, the Chilean Participation Group, the French Participation Group, Harvard-Smithsonian Center for Astrophysics, 
Instituto de Astrof\'isica de Canarias, The Johns Hopkins University, Kavli Institute for the Physics and Mathematics of the Universe (IPMU) / 
University of Tokyo, the Korean Participation Group, Lawrence Berkeley National Laboratory, 
Leibniz Institut f\"ur Astrophysik Potsdam (AIP),  
Max-Planck-Institut f\"ur Astronomie (MPIA Heidelberg), 
Max-Planck-Institut f\"ur Astrophysik (MPA Garching), 
Max-Planck-Institut f\"ur Extraterrestrische Physik (MPE), 
National Astronomical Observatories of China, New Mexico State University, 
New York University, University of Notre Dame, 
Observat\'ario Nacional / MCTI, The Ohio State University, 
Pennsylvania State University, Shanghai Astronomical Observatory, 
United Kingdom Participation Group,
Universidad Nacional Aut\'onoma de M\'exico, University of Arizona, 
University of Colorado Boulder, University of Oxford, University of Portsmouth, 
University of Utah, University of Virginia, University of Washington, University of Wisconsin, 
Vanderbilt University, and Yale University.


\begin{thebibliography}{}
\expandafter\ifx\csname natexlab\endcsname\relax\def\natexlab#1{#1}\fi

\bibitem[{{Anderson} {et~al.}(2010){Anderson}, {Covarrubias}, {James}, {Hamuy},
  \& {Habergham}}]{Anderson:2010aa}
{Anderson}, J.~P., {Covarrubias}, R.~A., {James}, P.~A., {Hamuy}, M., \&
  {Habergham}, S.~M. 2010, \mnras, 407, 2660

\bibitem[{{Anderson} \& {James}(2008)}]{Anderson:2008aa}
{Anderson}, J.~P., \& {James}, P.~A. 2008, \mnras, 390, 1527

\bibitem[{{Arcavi} {et~al.}(2010){Arcavi}, {Gal-Yam}, {Kasliwal}, {Quimby},
  {Ofek}, {Kulkarni}, {Nugent}, {Cenko}, {Bloom}, {Sullivan}, {Howell},
  {Poznanski}, {Filippenko}, {Law}, {Hook}, {J{\"o}nsson}, {Blake}, {Cooke},
  {Dekany}, {Rahmer}, {Hale}, {Smith}, {Zolkower}, {Velur}, {Walters},
  {Henning}, {Bui}, {McKenna}, \& {Jacobsen}}]{Arcavi:2010aa}
{Arcavi}, I., {Gal-Yam}, A., {Kasliwal}, M.~M., {et~al.} 2010, \apj, 721, 777

\bibitem[{{Arcavi} {et~al.}(2014){Arcavi}, {Gal-Yam}, {Sullivan}, {Pan},
  {Cenko}, {Horesh}, {Ofek}, {De Cia}, {Yan}, {Yang}, {Howell}, {Tal},
  {Kulkarni}, {Tendulkar}, {Tang}, {Xu}, {Sternberg}, {Cohen}, {Bloom},
  {Nugent}, {Kasliwal}, {Perley}, {Quimby}, {Miller}, {Theissen}, \&
  {Laher}}]{Arcavi:2014aa}
{Arcavi}, I., {Gal-Yam}, A., {Sullivan}, M., {et~al.} 2014, \apj, 793, 38

\bibitem[{{Asplund} {et~al.}(2009){Asplund}, {Grevesse}, {Sauval}, \&
  {Scott}}]{Asplund:2009fh}
{Asplund}, M., {Grevesse}, N., {Sauval}, A.~J., \& {Scott}, P. 2009, \araa, 47,
  481

\bibitem[{{Baldwin} {et~al.}(1981){Baldwin}, {Phillips}, \&
  {Terlevich}}]{Baldwin:1981aa}
{Baldwin}, J.~A., {Phillips}, M.~M., \& {Terlevich}, R. 1981, \pasp, 93, 5

\bibitem[{{Bietenholz} {et~al.}(2018){Bietenholz}, {Margutti}, {Alexander},
  {Argo}, {Bartel}, {Coppejans}, {Drout}, {Eftekhari}, {Guidorzi},
  {Milisavljevic}, \& {Terreran}}]{Bietenholz:2018aa}
{Bietenholz}, M., {Margutti}, R., {Alexander}, K., {et~al.} 2018, The
  Astronomer's Telegram, 11900

\bibitem[{{Bigiel} {et~al.}(2008){Bigiel}, {Leroy}, {Walter}, {Brinks}, {de
  Blok}, {Madore}, \& {Thornley}}]{Bigiel:2008rt}
{Bigiel}, F., {Leroy}, A., {Walter}, F., {et~al.} 2008, \aj, 136, 2846

\bibitem[{{Bolatto} {et~al.}(2013){Bolatto}, {Wolfire}, \&
  {Leroy}}]{Bolatto:2013rr}
{Bolatto}, A.~D., {Wolfire}, M., \& {Leroy}, A.~K. 2013, \araa, 51, 207

\bibitem[{{Catinella} {et~al.}(2018){Catinella}, {Saintonge}, {Janowiecki},
  {Cortese}, {Dav{\'e}}, {Lemonias}, {Cooper}, {Schiminovich}, {Hummels},
  {Fabello}, {Ger{\'e}b}, {Kilborn}, \& {Wang}}]{Catinella:2018aa}
{Catinella}, B., {Saintonge}, A., {Janowiecki}, S., {et~al.} 2018, \mnras, 476,
  875

\bibitem[{{Chabrier}(2003)}]{Chabrier:2003oe}
{Chabrier}, G. 2003, \pasp, 115, 763

\bibitem[{{Crowther}(2013)}]{Crowther:2013aa}
{Crowther}, P.~A. 2013, \mnras, 428, 1927

\bibitem[{{Denicol{\'o}} {et~al.}(2002){Denicol{\'o}}, {Terlevich}, \&
  {Terlevich}}]{Denicolo:2002aa}
{Denicol{\'o}}, G., {Terlevich}, R., \& {Terlevich}, E. 2002, \mnras, 330, 69

\bibitem[{{Drout} {et~al.}(2014){Drout}, {Chornock}, {Soderberg}, {Sanders},
  {McKinnon}, {Rest}, {Foley}, {Milisavljevic}, {Margutti}, {Berger},
  {Calkins}, {Fong}, {Gezari}, {Huber}, {Kankare}, {Kirshner}, {Leibler},
  {Lunnan}, {Mattila}, {Marion}, {Narayan}, {Riess}, {Roth}, {Scolnic},
  {Smartt}, {Tonry}, {Burgett}, {Chambers}, {Hodapp}, {Jedicke}, {Kaiser},
  {Magnier}, {Metcalfe}, {Morgan}, {Price}, \& {Waters}}]{Drout:2014aa}
{Drout}, M.~R., {Chornock}, R., {Soderberg}, A.~M., {et~al.} 2014, \apj, 794,
  23

\bibitem[{{French} {et~al.}(2016){French}, {Arcavi}, \&
  {Zabludoff}}]{French:2016aa}
{French}, K.~D., {Arcavi}, I., \& {Zabludoff}, A. 2016, \apjl, 818, L21

\bibitem[{{French} {et~al.}(2017){French}, {Arcavi}, \&
  {Zabludoff}}]{French:2017aa}
---. 2017, \apj, 835, 176

\bibitem[{{Galbany} {et~al.}(2017){Galbany}, {Mora}, {Gonz{\'a}lez-Gait{\'a}n},
  {Bolatto}, {Dannerbauer}, {L{\'o}pez-S{\'a}nchez}, {Maeda}, {P{\'e}rez},
  {P{\'e}rez-Torres}, {S{\'a}nchez}, {Wong}, {Badenes}, {Blitz}, {Marino},
  {Utomo}, \& {Van de Ven}}]{Galbany:2017aa}
{Galbany}, L., {Mora}, L., {Gonz{\'a}lez-Gait{\'a}n}, S., {et~al.} 2017,
  \mnras, 468, 628

\bibitem[{{Gezari} {et~al.}(2012){Gezari}, {Chornock}, {Rest}, {Huber},
  {Forster}, {Berger}, {Challis}, {Neill}, {Martin}, {Heckman}, {Lawrence},
  {Norman}, {Narayan}, {Foley}, {Marion}, {Scolnic}, {Chomiuk}, {Soderberg},
  {Smith}, {Kirshner}, {Riess}, {Smartt}, {Stubbs}, {Tonry}, {Wood-Vasey},
  {Burgett}, {Chambers}, {Grav}, {Heasley}, {Kaiser}, {Kudritzki}, {Magnier},
  {Morgan}, \& {Price}}]{Gezari:2012aa}
{Gezari}, S., {Chornock}, R., {Rest}, A., {et~al.} 2012, \nat, 485, 217

\bibitem[{{Hakobyan} {et~al.}(2014){Hakobyan}, {Nazaryan}, {Adibekyan},
  {Petrosian}, {Aramyan}, {Kunth}, {Mamon}, {de Lapparent}, {Bertin}, {Gomes},
  \& {Turatto}}]{Hakobyan:2014aa}
{Hakobyan}, A.~A., {Nazaryan}, T.~A., {Adibekyan}, V.~Z., {et~al.} 2014,
  \mnras, 444, 2428

\bibitem[{{Hatsukade} {et~al.}(2019){Hatsukade}, {Hashimoto}, {Kohno},
  {Nakanishi}, {Ohta}, {Niino}, {Tamura}, \& {T{\'o}th}}]{Hatsukade:2019aa}
{Hatsukade}, B., {Hashimoto}, T., {Kohno}, K., {et~al.} 2019, \apj, 876, 91

\bibitem[{{Ho} {et~al.}(2019){Ho}, {Phinney}, {Ravi}, {Kulkarni}, {Petitpas}, {Emonts}, {Bhalerao}, {Blundell}, {Cenko}, {Dobie}, {Howie}, {Kamraj}, {Kasliwal}, {Murphy}, {Perley}, {Sridharan}, \& {Yoon}}]{Ho:2019aa}
{Ho}, A.~Y.~Q., {Phinney}, E.~S., {Ravi}, V., {et~al.} 2019, \apj, 871, 73

\bibitem[{{Japelj} {et~al.}(2018){Japelj}, {Vergani}, {Salvaterra}, {Renzo},
  {Zapartas}, {de Mink}, {Kaper}, \& {Zibetti}}]{Japelj:2018aa}
{Japelj}, J., {Vergani}, S.~D., {Salvaterra}, R., {et~al.} 2018, \aap, 617,
  A105

\bibitem[{{Komossa} \& {Bade}(1999)}]{Komossa:1999aa}
{Komossa}, S., \& {Bade}, N. 1999, \aap, 343, 775

\bibitem[{{Kuin} {et~al.}(2019){Kuin}, {Wu}, {Oates}, {Lien}, {Emery},
  {Kennea}, {de Pasquale}, {Han}, {Brown}, {Tohuvavohu}, {Breeveld}, {Burrows},
  {Cenko}, {Campana}, {Levan}, {Markwardt}, {Osborne}, {Page}, {Page},
  {Sbarufatti}, {Siegel}, \& {Troja}}]{Kuin:2019aa}
{Kuin}, N.~P.~M., {Wu}, K., {Oates}, S., {et~al.} 2019, \mnras,
  arXiv:1808.08492

\bibitem[{{Kuncarayakti} {et~al.}(2018){Kuncarayakti}, {Anderson}, {Galbany},
  {Maeda}, {Hamuy}, {Aldering}, {Arimoto}, {Doi}, {Morokuma}, \&
  {Usuda}}]{Kuncarayakti:2018aa}
{Kuncarayakti}, H., {Anderson}, J.~P., {Galbany}, L., {et~al.} 2018, \aap, 613,
  A35

\bibitem[{{Lazzati} {et~al.}(2012){Lazzati}, {Morsony}, {Blackwell}, \&
  {Begelman}}]{Lazzati:2012aa}
{Lazzati}, D., {Morsony}, B.~J., {Blackwell}, C.~H., \& {Begelman}, M.~C. 2012,
  \apj, 750, 68

\bibitem[{{Lin} {et~al.}(2018){Lin}, {Strader}, {Carrasco}, {Page},
  {Romanowsky}, {Homan}, {Irwin}, {Remillard}, {Godet}, {Webb}, {Baumgardt},
  {Wijnands}, {Barret}, {Duc}, {Brodie}, \& {Gwyn}}]{Lin:2018aa}
{Lin}, D., {Strader}, J., {Carrasco}, E.~R., {et~al.} 2018, Nature Astronomy,
  2, 656

\bibitem[{{Margutti} {et~al.}(2019){Margutti}, {Metzger}, {Chornock}, {Vurm}, {Roth}, {Grefenstette}, {Savchenko}, {Cartier}, {Steiner}, {Terreran}, {Margalit}, {Migliori}, {Milisavljevic}, {Alexander}, {Bietenholz}, {Blanchard}, {Bozzo}, {Brethauer}, {Chilingarian}, {Coppejans}, {Ducci}, {Ferrigno}, {Fong}, {G{\"o}tz}, {Guidorzi}, {Hajela}, {Hurley}, {Kuulkers}, {Laurent}, {Mereghetti}, {Nicholl}, {Patnaude}, {Ubertini}, {Banovetz}, {Bartel}, {Berger}, {Coughlin}, {Eftekhari}, {Frederiks}, {Kozlova}, {Laskar}, {Svinkin}, {Drout}, {MacFadyen}, \& {Paterson}}]{Margutti:2019aa}
{Margutti}, R., {Metzger}, B.~D., {Chornock}, R., {et~al.} 2019, \apj, 872, 18

\bibitem[{{McGaugh}(1991)}]{McGaugh:1991ee}
{McGaugh}, S.~S. 1991, \apj, 380, 140

\bibitem[{{McMullin} {et~al.}(2007){McMullin}, {Waters}, {Schiebel}, {Young},
  \& {Golap}}]{McMullin:2007aa}
{McMullin}, J.~P., {Waters}, B., {Schiebel}, D., {Young}, W., \& {Golap}, K.
  2007, in Astronomical Society of the Pacific Conference Series, Vol. 376,
  Astronomical Data Analysis Software and Systems XVI, ed. R.~A. {Shaw},
  F.~{Hill}, \& D.~J. {Bell}, 127

\bibitem[{{Micha{\l}owski} {et~al.}(2019){Micha{\l}owski}, {Kamphuis},
  {Hjorth}, {Kann}, {de Ugarte Postigo}, {Galbany}, {Fynbo}, {Ghosh}, {Hunt},
  {Kuncarayakti}, {Le Floc'h}, {Le{\'s}niewska}, {Misra}, {Nicuesa Guelbenzu},
  {Palazzi}, {Rasmussen}, {Resmi}, {Rossi}, {Savaglio}, {Schady}, {Schulze},
  {Th{\"o}ne}, {Watson}, {J{\'o}zsa}, {Serra}, \&
  {Smirnov}}]{Michaowski:2019aa}
{Micha{\l}owski}, M.~J., {Kamphuis}, P., {Hjorth}, J., {et~al.} 2019, arXiv
  e-prints, arXiv:1902.10144

\bibitem[{{Miller} {et~al.}(2015){Miller}, {Kaastra}, {Miller}, {Reynolds},
  {Brown}, {Cenko}, {Drake}, {Gezari}, {Guillochon}, {Gultekin}, {Irwin},
  {Levan}, {Maitra}, {Maksym}, {Mushotzky}, {O'Brien}, {Paerels}, {de Plaa},
  {Ramirez-Ruiz}, {Strohmayer}, \& {Tanvir}}]{Miller:2015aa}
{Miller}, J.~M., {Kaastra}, J.~S., {Miller}, M.~C., {et~al.} 2015, \nat, 526,
  542

\bibitem[{{Modjaz} {et~al.}(2008){Modjaz}, {Kewley}, {Kirshner}, {Stanek},
  {Challis}, {Garnavich}, {Greene}, {Kelly}, \& {Prieto}}]{Modjaz:2008aa}
{Modjaz}, M., {Kewley}, L., {Kirshner}, R.~P., {et~al.} 2008, \aj, 135, 1136

\bibitem[{{Modjaz} {et~al.}(2019){Modjaz}, {Bianco}, {Siwek}, {Huang},
  {Perley}, {Fierroz}, {Liu}, {Arcavi}, {Gal-Yam}, {Blagorodnova}, {Cenko},
  {Filippenko}, {Kasliwal}, {Kulkarni}, {Schulze}, {Taggart}, \&
  {Zhen}}]{Modjaz:2019aa}
{Modjaz}, M., {Bianco}, F.~B., {Siwek}, M., {et~al.} 2019, arXiv e-prints,
  arXiv:1901.00872

\bibitem[{{Perley} {et~al.}(2019){Perley}, {Mazzali}, {Yan}, {Cenko}, {Gezari},
  {Taggart}, {Blagorodnova}, {Fremling}, {Mockler}, {Singh}, {Tominaga},
  {Tanaka}, {Watson}, {Ahumada}, {Anupama}, {Ashall}, {Becerra}, {Bersier},
  {Bhalerao}, {Bloom}, {Butler}, {Copperwheat}, {Coughlin}, {De}, {Drake},
  {Duev}, {Frederick}, {Gonz{\'a}lez}, {Goobar}, {Heida}, {Ho}, {Horst},
  {Hung}, {Itoh}, {Jencson}, {Kasliwal}, {Kawai}, {Khanam}, {Kulkarni},
  {Kumar}, {Kumar}, {Kutyrev}, {Lee}, {Maeda}, {Mahabal}, {Murata}, {Neill},
  {Ngeow}, {Penprase}, {Pian}, {Quimby}, {Ramirez-Ruiz}, {Richer},
  {Rom{\'a}n-Z{\'u}{\~n}iga}, {Sahu}, {Srivastav}, {Socia}, {Sollerman},
  {Tachibana}, {Taddia}, {Tinyanont}, {Troja}, {Ward}, {Wee}, \&
  {Yu}}]{Perley:2019aa}
{Perley}, D.~A., {Mazzali}, P.~A., {Yan}, L., {et~al.} 2019, \mnras, 484, 1031

\bibitem[{{Petry} \& {CASA Development Team}(2012)}]{Petry:2012aa}
{Petry}, D., \& {CASA Development Team}. 2012, in Astronomical Society of the
  Pacific Conference Series, Vol. 461, Astronomical Data Analysis Software and
  Systems XXI, ed. P.~{Ballester}, D.~{Egret}, \& N.~P.~F. {Lorente}, 849

\bibitem[{{Pettini} \& {Pagel}(2004)}]{Pettini:2004jl}
{Pettini}, M., \& {Pagel}, B.~E.~J. 2004, \mnras, 348, L59

\bibitem[{{Pilyugin} \& {Thuan}(2005)}]{Pilyugin:2005os}
{Pilyugin}, L.~S., \& {Thuan}, T.~X. 2005, \apj, 631, 231

\bibitem[{{Prentice} {et~al.}(2018){Prentice}, {Maguire}, {Smartt}, {Magee},
  {Schady}, {Sim}, {Chen}, {Clark}, {Colin}, {Fulton}, {McBrien}, {O'Neill},
  {Smith}, {Ashall}, {Chambers}, {Denneau}, {Flewelling}, {Heinze}, {Holoien},
  {Huber}, {Kochanek}, {Mazzali}, {Prieto}, {Rest}, {Shappee}, {Stalder},
  {Stanek}, {Stritzinger}, {Thompson}, \& {Tonry}}]{Prentice:2018aa}
{Prentice}, S.~J., {Maguire}, K., {Smartt}, S.~J., {et~al.} 2018, \apjl, 865,
  L3

\bibitem[{{Pursiainen} {et~al.}(2018){Pursiainen}, {Childress}, {Smith},
  {Prajs}, {Sullivan}, {Davis}, {Foley}, {Asorey}, {Calcino}, {Carollo},
  {Curtin}, {D'Andrea}, {Glazebrook}, {Gutierrez}, {Hinton}, {Hoormann},
  {Inserra}, {Kessler}, {King}, {Kuehn}, {Lewis}, {Lidman}, {Macaulay},
  {M{\"o}ller}, {Nichol}, {Sako}, {Sommer}, {Swann}, {Tucker}, {Uddin},
  {Wiseman}, {Zhang}, {Abbott}, {Abdalla}, {Allam}, {Annis}, {Avila}, {Brooks},
  {Buckley-Geer}, {Burke}, {Carnero Rosell}, {Carrasco Kind}, {Carretero},
  {Castander}, {Cunha}, {Davis}, {De Vicente}, {Diehl}, {Doel}, {Eifler},
  {Flaugher}, {Fosalba}, {Frieman}, {Garc{\'{\i}}a-Bellido}, {Gruen},
  {Gruendl}, {Gutierrez}, {Hartley}, {Hollowood}, {Honscheid}, {James},
  {Jeltema}, {Kuropatkin}, {Li}, {Lima}, {Maia}, {Martini}, {Menanteau},
  {Ogando}, {Plazas}, {Roodman}, {Sanchez}, {Scarpine}, {Schindler}, {Smith},
  {Soares-Santos}, {Sobreira}, {Suchyta}, {Swanson}, {Tarle}, {Tucker}, \&
  {Walker}}]{Pursiainen:2018aa}
{Pursiainen}, M., {Childress}, M., {Smith}, M., {et~al.} 2018, \mnras, 481, 894

\bibitem[{{Rana} \& {Basu}(1992)}]{Rana:1992aa}
{Rana}, N.~C., \& {Basu}, S. 1992, \aap, 265, 499

\bibitem[{{Rest} {et~al.}(2018){Rest}, {Garnavich}, {Khatami}, {Kasen},
  {Tucker}, {Shaya}, {Olling}, {Mushotzky}, {Zenteno}, {Margheim},
  {Strampelli}, {James}, {Smith}, {F{\"o}rster}, \& {Villar}}]{Rest:2018aa}
{Rest}, A., {Garnavich}, P.~M., {Khatami}, D., {et~al.} 2018, Nature Astronomy,
  2, 307

\bibitem[{{Roychowdhury} {et~al.}(2019){Roychowdhury}, {Arabsalmani}, \&
  {Kanekar}}]{Roychowdhury:2019aa}
{Roychowdhury}, S., {Arabsalmani}, M., \& {Kanekar}, N. 2019, \mnras, 485, L93

\bibitem[{{Saintonge} {et~al.}(2017){Saintonge}, {Catinella}, {Tacconi},
  {Kauffmann}, {Genzel}, {Cortese}, {Dav{\'e}}, {Fletcher},
  {Graci{\'a}-Carpio}, {Kramer}, {Heckman}, {Janowiecki}, {Lutz}, {Rosario},
  {Schiminovich}, {Schuster}, {Wang}, {Wuyts}, {Borthakur}, {Lamperti}, \&
  {Roberts-Borsani}}]{Saintonge:2017aa}
{Saintonge}, A., {Catinella}, B., {Tacconi}, L.~J., {et~al.} 2017, \apjs, 233,
  22

\bibitem[{{Speagle} {et~al.}(2014){Speagle}, {Steinhardt}, {Capak}, \&
  {Silverman}}]{Speagle:2014aa}
{Speagle}, J.~S., {Steinhardt}, C.~L., {Capak}, P.~L., \& {Silverman}, J.~D.
  2014, \apjs, 214, 15

\bibitem[{{Tremonti} {et~al.}(2004){Tremonti}, {Heckman}, {Kauffmann},
  {Brinchmann}, {Charlot}, {White}, {Seibert}, {Peng}, {Schlegel}, {Uomoto},
  {Fukugita}, \& {Brinkmann}}]{Tremonti:2004rf}
{Tremonti}, C.~A., {Heckman}, T.~M., {Kauffmann}, G., {et~al.} 2004, \apj, 613,
  898

\bibitem[{{Zaritsky} {et~al.}(1994){Zaritsky}, {Kennicutt}, \&
  {Huchra}}]{Zaritsky:1994zt}
{Zaritsky}, D., {Kennicutt}, Jr., R.~C., \& {Huchra}, J.~P. 1994, \apj, 420, 87

\end{thebibliography}



\end{document}